\pdfoutput=1
\documentclass[journal=nalefd,manuscript=article]{achemso}
\usepackage[utf8]{inputenc}
\usepackage{mciteplus}
\usepackage[T1]{fontenc} % Use modern font encodings
\usepackage{amsmath}

\author{Mason J. Gray}
\author{Josef Freudenstein}
\affiliation{Department of Physics, Boston College, Chestnut Hill, MA. 02467, United States}
\author{Shu Yang F. Zhao}
\affiliation{Department of Physics, Harvard University, Cambridge, Massachusetts 02138, United States}
\author{Ryan O'Connor}
\author{Samuel Jenkins}
\author{Narendra Kumar}
\author{Marcel Hoek}
\author{Abigail Kopec}
\affiliation{Department of Physics, Boston College, Chestnut Hill, MA. 02467, United States}
\author{Soonsang Huh}
\affiliation{Department of Physics and Astronomy, Seoul National University (SNU), Seoul 08826, Republic of Korea}
\author{Takashi Taniguchi}
\author{Kenji Watanabe}
\affiliation{National Institute for Materials Science, 1-1 Namiki, Tsukuba 306-0044, Japan}
\author{Ruidan Zhong}
\affiliation{Condensed Matter Physics and Materials Science Department, Brookhaven National Laboratory, Upton, New York 11973, United States }
\author{Changyoung Kim}
\affiliation{Department of Physics and Astronomy, Seoul National University (SNU), Seoul 08826, Republic of Korea}
\author{G. D. Gu}
\affiliation{Condensed Matter Physics and Materials Science Department, Brookhaven National Laboratory, Upton, New York 11973, United States }
\author{K. S. Burch}
\affiliation{Department of Physics, Boston College, Chestnut Hill, MA. 02467, United States}
\email{ks.burch@bc.com}

\title{Evidence for Helical Hinge Zero Modes in an Fe-Based Superconductor}

\begin{document}

\begin{abstract}
    Combining topology and superconductivity provides a powerful tool for investigating fundamental physics as well as a route to fault-tolerant quantum computing. There is mounting evidence that the Fe-Based superconductor FeTe$_{0.55}$Se$_{0.45}$ (FTS) may also be topologically non-trivial. Should the superconducting order be s$^{\pm}$, then FTS could be a higher order topological superconductor with Helical Hinge Zero Modes (HHZM). To test the presence of these modes we've fabricated normal-metal/superconductor junctions on different surfaces via 2D atomic crystal heterostructures. As expected, junctions in contact with the hinge reveal a sharp zero-bias anomaly that is absent when tunneling purely into the c-axis. Additionally, the shape and suppression with temperature are consistent with highly coherent modes along the hinge and are  incongruous with other origins of zero bias anomalies. Furthermore, additional measurements with soft-point contacts in bulk samples with various Fe interstitial contents demonstrate the intrinsic nature of the observed mode. Thus we provide evidence that FTS is indeed a higher order topological superconductor. 
\end{abstract}

\textbf{Keywords:} Higher-order topology, 2D superconductor, Hinge modes, Andreev Reflection

\section{Introduction}
New particles can be a convincing signature of emergent phases of matter, from spinons in quantum spin liquids\cite{Balents2010} to the Fermi arcs of Weyl semimetals\cite{Armitage2017,Zhang2019}. Beyond potentially indicating a broken symmetry or topological invariant, they can be put to use in future topological quantum computers\cite{Nayak2008}. Until recently it was believed the non-trivial topology of the bulk would lead to new states in one lower dimension at the boundary with a system of differing topology. However, higher order topological insulators (HOTI) have been realized\cite{Schindler2018,Ni2018,Xue2019,Song2017,Langbehn2017,Benalcazar2017}, where the resulting boundary modes exist only at the intersection of two or more edges, producing 1D hinge or 0D bound states. One route to creating these higher order states is through the combination of a topological insulator and a superconductor with anisotropic pairing\cite{Wang2018Corner,DasSarma2018,Zhongbo2018,Ghorashi2019}. Usually, this is done by combining two separate materials and inducing superconductivity into the TI via proximity\cite{Zareapour2012,Albrecht2016,Gazibegovic2017,Kurter2018,Tanaka2012}. However, this method requires long coherence lengths and extremely clean interfaces, making experimental realization of devices quite difficult. For studying HOTI, as well as the combination of strong correlations and topology, the material FeTe$_{0.55}$Se$_{0.45}$ (FTS) may be ideal, as it is a bulk, high-temperature superconductor with anisotropic pairing that also hosts topologically non-trivial surface states\cite{Zhang2018,Wang2015,Wang2018}.
\begin{figure}[H]
    \centering
    \includegraphics[width=\textwidth]{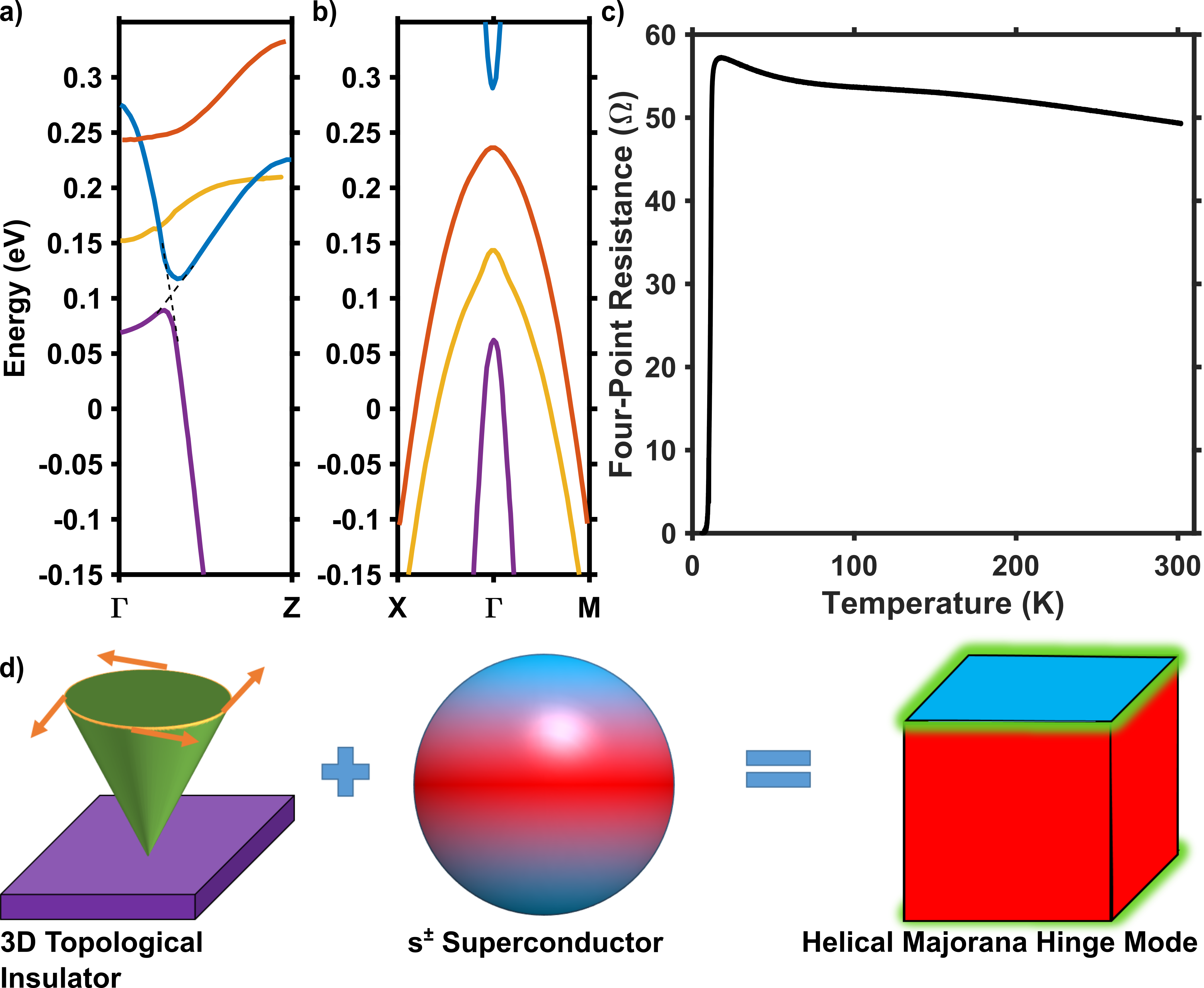}
    \caption{a) Theoretical band structure of FeTe$_{0.55}$Se$_{0.45}$ along the $\Gamma$-Z and (b) the X-$\Gamma$-M cuts\cite{Wang2015}. The $p_{z}$ orbital of the chalcogenide is shown in blue, crossing the three d-orbitals, resulting in two Dirac points and topological, spin-orbit gap. c) Resistance vs. Temperature graph for an exfoliated flake of FTS, showing a clear superconducting transition around 10K. d) Diagram showing the ingredients needed for a Helical Majorana Hinge Mode}
    \label{HingeTheory}
\end{figure}
\par
FTS is part of the FeTe$_{1-x}$Se$_{x}$ family of Fe-based superconductors, which ranges from an antiferromagnet in FeTe to a bulk superconductor in FeSe\cite{Liu2010}. These generally have the same Fermiology as the other Fe-based superconductors in that there are hole pockets at the $\Gamma$-point and electron pockets at the M-points\cite{Homes2015,Hanaguri2010,Miao2012,Okazaki2012,Zhang2018}. The relative strengths of the interband vs intraband scattering in principle should determine the superconducting symmetry, however, there is a complex interplay between the spin-fluctuation exchange, intraband Coulomb repulsion, and the doping level that all contribute to the symmetry of the superconducting order parameter\cite{Kreisel2016,Chubukov2012}. Indeed, experiments performed on FeTe$_{0.55}$Se$_{0.45}$ find no evidence for a node with strong signatures of s$^{\pm}$ order,\cite{Hanaguri2010,Miao2012,Zeng2010} while experiments on other alloys suggest nodal s$^{\pm}$, anisotropic s-wave, and even p-wave\cite{Michioka2010,Serafin2010,Bendele2010,Kim2010,Okazaki2012}. Interestingly, tuning away from FeSe leads to enhanced spin-orbit coupling and bandwidth. As a result, the p-orbital is shifted down in energy, crossing the d-orbitals with opposite parity along the $\Gamma$ to $Z$ direction (See Figure \ref{HingeTheory}a and b). The first two crossings are protected by crystalline-symmetry resulting in bulk Dirac states above the Fermi energy. However, the lowest energy crossing is avoided resulting in a spin-orbit coupled gap, resembling those typically found in topological insulators\cite{Wang2015,Chen178}. While the Fermi level falls into this gap, the original hole and electron Fermi surfaces at $\Gamma$ and $M$, respectively, are retained\cite{Zhang2018,Wang2015}. ARPES measurements have observed the resulting spin-momentum locked surface states, as well as their gaping out in the superconducting state\cite{Zhang2018,Zhang20192}. Additionally, there is evidence from STM that this results in apparent Majorana zero-modes inside magnetic vortices\cite{Wang2018,Dong-LaiFeng2018,Machida2018}.
\par
Recent theoretical work on FTS has suggested that the combination of an s$^{\pm}$ order parameter and topological surface states could give rise to higher order topological superconductivity\cite{DasSarma2018}. In short, the changing superconducting phase causes the surface states to gap out anisotropically. Depending on the relative strength of the isotropic versus the anisotropic term, this could lead to the [001] and the [100] or [010] face having superconducting order parameters with opposite phase. As shown in Figure \ref{HingeTheory}d), this is predicted to produce a pair of 1D Helical Majorana Hinge Modes emerging at the 1D interface of the top/side surfaces\cite{DasSarma2018}. Whether or not the modes we observe are indeed Majorana modes, the appearance of HHZM requires both s$^{\pm}$ superconductivity as well as strong 3D TI surface states. Thus observing Helical Hinge Zero Modes in FTS would provide strong evidence that it is an s$^{\pm}$ topological superconductor.
\par
\begin{figure}[H]
    \centering
    \includegraphics[width=\textwidth]{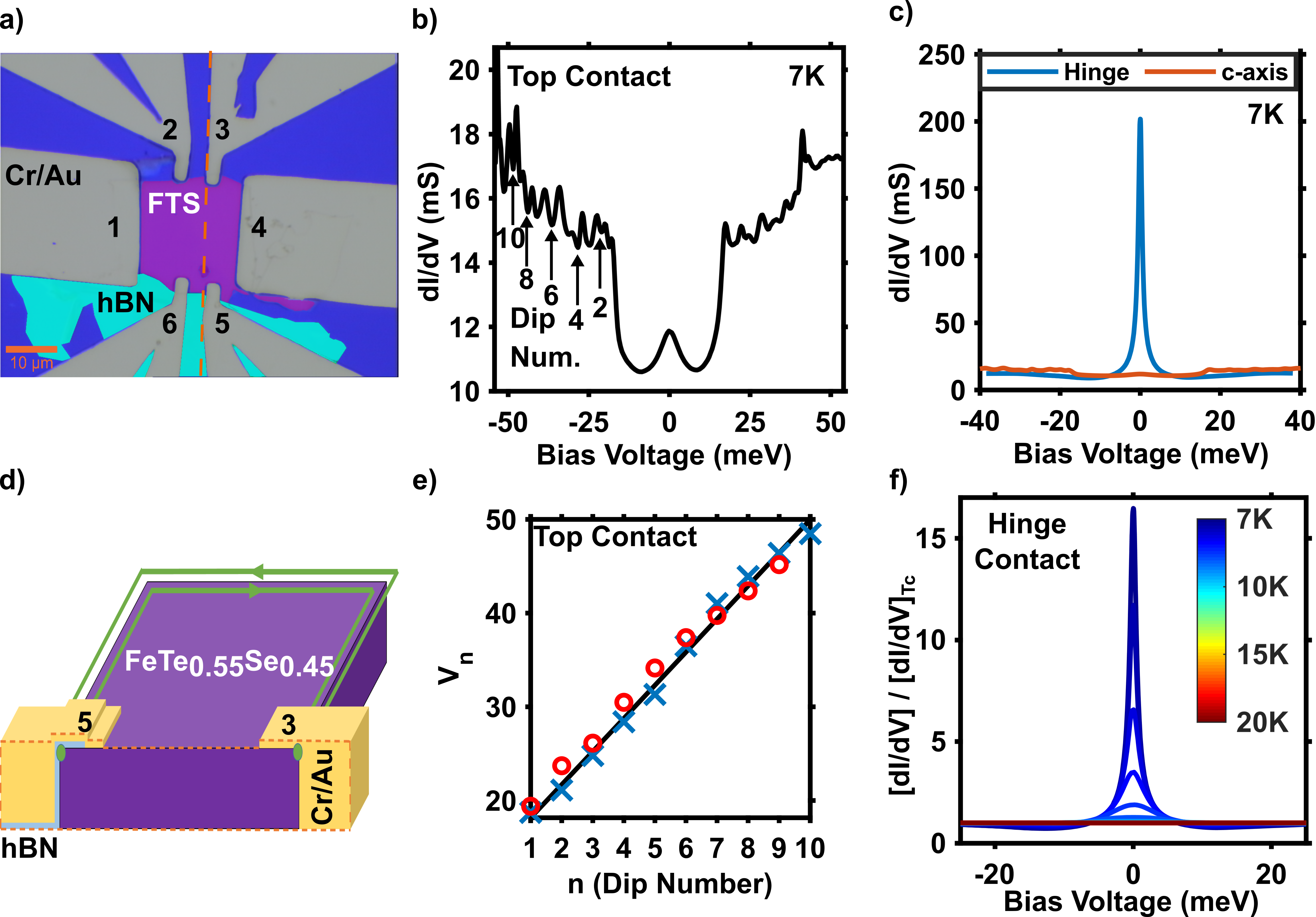}
    \caption{a) False color image of the exfoliated device; numbers denote contacts used. b) $\frac{dI}{dV}$ vs DC Bias voltage for contact 5 at 7 K. c)$\frac{dI}{dV}$ vs DC Bias voltage for contact 3 at 7 K. d) Depiction of contact geometry for top only (5) and hinge (3) contacts. e) Dip number vs. Voltage for c-axis only contacts. The black line is a fit to McMillan-Rowell Oscillations which follow the equation, $\Delta V = n\times\frac{hv_{F}}{4ed_{s}}$. Blue and red points are experimental data extracted from the positive and negative bias voltages respectively. f) Temperature dependence of differential conductance for various temperatures.}
    \label{MainResults}
\end{figure}

To search for the HHZM it is tempting to rely on methods previously exploited to reveal the unconventional nature of the cuprates\cite{Deutscher2005}. Specifically, normal-metal/superconductor junctions demonstrated Andreev Bound States resulting from the d-wave order only on [110] surfaces\cite{Tanaka2003,Sinha1998,Greene1999,Tanaka2012}. In the case of FTS, this approach is more challenging as one must tunnel into the hinge between [001] and [010] and the modes are nominally charge neutral, thus requiring an Andreev process to be observed\cite{zhang2017quantum}.  To achieve this, we created 2D atomic crystal heterostructures with thick hBN covering half of the FTS. By draping contacts over the side of the FTS or atop the hBN we can separately probe conductance into the hinge from the c-axis. As expected for modes protected from back-scattering, we find a cusp-like zero-bias peak only on the hinge contacts that is absent from the c-axis junctions. The mode is well-described by a Lorentzian, consistent with other studies on one-dimensional zero-energy bound states\cite{Setiawan2017}. Confirmation that the mode does not result from our fabrication method or defect density is provided by soft-point contact measurements on facets of various bulk crystals (See Supplemental Fig S3). Taken together these data strongly suggest the presence of the HHZM in FTS resulting from its higher order topological nature and the presence of s$^{\pm}$ superconductivity\cite{Park:2010wo,Tanaka2012}. 
The helical hinge zero mode in FTS should only exist in the superconducting state. As such we expect a sharp zero-bias conductance feature below T$_{c}$ on the hinges between the [001] and side surfaces as compared to purely on the [001] face. Alternatively, Majorana zero modes on the hinge should give quantized conductance, revealed through nearly perfect Andreev reflection.\cite{DasSarma2018} However, as discussed later, observing this quantized conductance may be challenging as the coherence length in FTS is $\approx 3 nm$\cite{Kim2010,Bendele2010}. To test this we used 2D atomic crystal heterostructures to simultaneously fabricate Normal Metal/Superconductor (NS) low-barrier junctions on various crystal facets (See Figure \ref{MainResults}a and \ref{MainResults}d). The first type of NS junction is a standard lithographically-defined contact that drapes over the edge of the exfoliated flake. This contact will form a junction with the [001] and [100] surfaces as well as the hinge between them. The second type of contact is fabricated by first transferring hexagonal Boron Nitride (hBN) over half of the FTS flake, insulating the side and edge from electrical contact. We then drape a contact over the side of the hBN, forming a junction primarily on the [001] face (See depiction of the side view in Fig \ref{MainResults}d). The entire fabrication process, from exfoliation to device, is performed in an inert argon atmosphere or vacuum. Patterns for mesoscale contacts were defined using standard photolithography techniques and our Heidelberg $\mu$PG101 direct-write lithography system. Contact areas are then cleaned with an argon plasma at high vacuum immediately before thermal deposition of 5nm of Cr then 45nm of Au. Full fabrication details can be found in the Supplementary.
\par
\section{Results and Discussion}
We first established that our control contacts are only tunneling into the c-axis by studying their base temperature differential conductance. Specifically, we sourced current between a top contact (5 or 6 in Fig. \ref{MainResults}a) to one of the current leads (\#1 or \#4), while measuring the resulting voltage between the same top contact and the other current contact. This three-point experiment ensures the conductance results primarily from the interface of the top contact. As shown in Fig \ref{MainResults}b, we observe a small zero bias conductance peak that is $\approx 20 \%$ higher than the background. The shape and height are consistent with previous point contact Andreev reflection measurements along the c-axis of FeTe$_{0.55}$Se$_{0.45}$,\cite{Daghero_2014} and confirms the contacts are in the low-bias, Andreev regime. We note these previous works were performed at temperatures below our base temperature, and as such could resolve the rather small gap. At higher bias, we observe an enhancement in the conductance  at $|V|\geq 20~meV$, consistent with spin-orbit induced gap. Above this value, we observe a series of conductance dips that are fully consistent with McMillan-Rowell Oscillations (MRO)\cite{2004PhyC..408..618C,2004PhRvB..69m2507S}. These MRO result from Fabry-Perot like interference of quasiparticles in the normal layer undergoing AR at the interface and reflecting off the back surface of the metal. The MRO are linearly spaced by voltages\cite{2004PhyC..408..618C} defined by the equation $\Delta(V) = n\cdot\frac{ev_{F}}{hd}$ where n is the dip number, $v_{F}$ is the Fermi velocity at the contact, and $d$ is the thickness of the metal which we set to 50~nm (See Figure 2e). From this fit, we extract a renormalized Fermi velocity of approximately $1.7\times10^5 m/s$.  We note that similar behavior was observed if the current/Voltage was reversed between contacts \#1 \& \#4, we measure from contact \#6, or measuring between contacts \#6 and \#5 exclusively (see Supplemental Fig S4a). This shows the robustness of these results and combined with the detailed spectra, confirm the contacts over the hBN are Andreev tunneling only into the c-axis. 

Next, we turn to the spectra measured in an identical manner, but with the hinge contact (\#3 in Fig. \ref{MainResults}a). Since the normal-state and high bias resistance of the hinge contact is nearly identical to the control contact we expect the spectra to be similar. However, as shown in Fig. \ref{MainResults}c the zero-bias conductance in the hinge contact is quite distinct from the response observed in the control contact and previous point contact experiments. Specifically, we observe a cusp-like zero-bias conductance peak (ZBCP) in the hinge contact that reaches a value 17-times higher than the high bias or $T\approx T_{c}$ conductance. This rather large enhancement is also likely responsible for the absence of a clear observation of the gap, which would be far smaller. These results provide strong evidence for a zero mode that only exists on the hinge. The "cusp-like" shape and magnitude of the peak could result from an Andreev Bound State (ABS)\cite{Deutscher2005,Sinha1998,Greene1999}, however, this requires either a node in the superconducting gap or time-reversal symmetry breaking,\cite{Yakovenko2002,TanakaTopSym2012} neither of which has been detected in FeTe$_{0.55}$Se$_{0.45}$\cite{Serafin2010,Zeng2010,Bendele2010,Okazaki2012,Miao2012,Kim2010,Hanaguri2010}. As discussed later, direct evidence against the ABS interpretation is provided by the dependence of the peak on temperature, and near independence on the contact's type (planar, point contact) or material (Ag, Au, Bi$_{2}$Te$_{2}$Se$_{1}$). Interestingly, this behavior is also inconsistent with previous observations of standard Andreev Reflection(AR)\cite{Tanaka2003}, Coherent Andreev Reflection (CAR)\cite{Klapwijk1992}, the Kondo Effect\cite{Sasaki2000,Samokhin2001}, and Joule heating\cite{Naidyuk2018}. 

\begin{figure}[H]
    \centering
    \includegraphics[width=\textwidth]{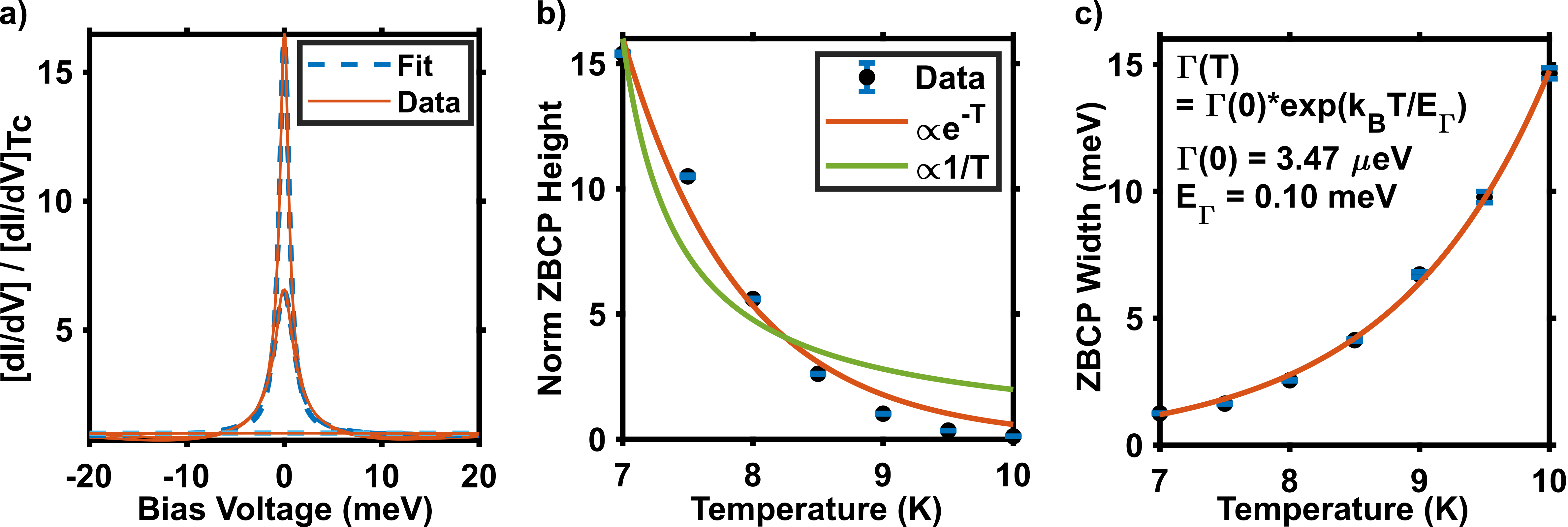}
    \caption{a) dI/dV versus voltage normalized to the spectra taken at T$_{c}$ (solid line) with a Lorentzian fit (dashed line), for $T=7K$, $9K$, and $15K$. b) and c) ZBCP heights and widths, respectively, extracted from the Lorentzian fit versus temperature. The exponential temperature dependence (orange lines) is at odds with a normal Andreev bound state that follows a $1/T$ dependence. The small energy scale of the exponential may result from the reduced superconducting gap on the side surfaces. While the rather small width at zero temperature is consistent with a topologically protected 1D mode.}
    \label{DataAnalysis}
\end{figure}

To ensure the zero bias conductance peak emerges at T$_{c}$ and is not the result of an ABS, we directly analyzed its temperature dependence by fitting the data with a Lorentzian line shape. This is based on recent theoretical studies on one-dimensional superconducting wires showing that both Majorana Zero Modes and ABS produce a Lorentzian differential conductance spectra\cite{Setiawan2017}. While this may not be the correct model for our case, to the best of our knowledge there are no calculations for the conductance spectra expected from hinge modes in a higher order topological superconductor. Nonetheless, the differential conductance spectra are generally well described by a Lorentzian (see Fig \ref{DataAnalysis}a). The temperature dependence of the height and width of the peak determined by the fits for the data presented in Fig. \ref{MainResults}f are shown in Fig. \ref{DataAnalysis}b \& c, respectively. These data provide direct evidence for the connection to the bulk superconductivity, though are inconsistent with an ABS. Indeed, we find that as the temperature is raised, the height of the ZBCP decreases exponentially until it is completely quenched at $T_{c}$ (see Fig\ref{MainResults}a and Fig\ref{DataAnalysis}b), where we define $T_{C}$ as the temperature for which $\frac{dR}{dT}$ passes through zero. While lower temperature data are required to determine the exact functional form, it is clear from Fig. \ref{DataAnalysis}b \& c that the mode is substantially different from the $1/T$ behavior typically expected from an ABS. Furthermore, we found a similar shape and temperature dependence in contacts of various barrier height, also inconsistent with standard Andreev reflection.\cite{BTK,Tanaka2012,Lofwander2001}

Similar to the height of the peak, we find the width of the zero bias conductance peak grows exponentially with temperature (see Fig. \ref{DataAnalysis}). Interestingly the energy scale governing the peak height ($E_{H}\approx 0.08~meV$) and the width ($E_{\Gamma}\approx 0.1~meV$) are quite close. We note that comparable results were obtained from other contacts revealing the hinge mode. Nonetheless, the energy scales governing the temperature dependence of the mode are far smaller than either the superconducting gap of the bulk or the surface states.\cite{Zhang2018} However, to the best of our knowledge, the size of the superconducting gap on the side surface has not been measured. As such we speculate this small apparent energy scale results from a much weaker proximity effect on the [010] and [100] surface states. Interestingly, extrapolating the width of the zero bias peak to zero temperature suggests an extremely narrow mode ($\approx 3.5~\mu eV$). While further studies at lower temperatures are required to confirm this extrapolation and the specific shape of the mode, if correct it points to the highly coherent nature of the excitation. As such the temperature dependence is consistent with our expectations for topologically protected 1D modes. 
\par
\begin{figure}[H]
    \centering
    \includegraphics[width=\textwidth]{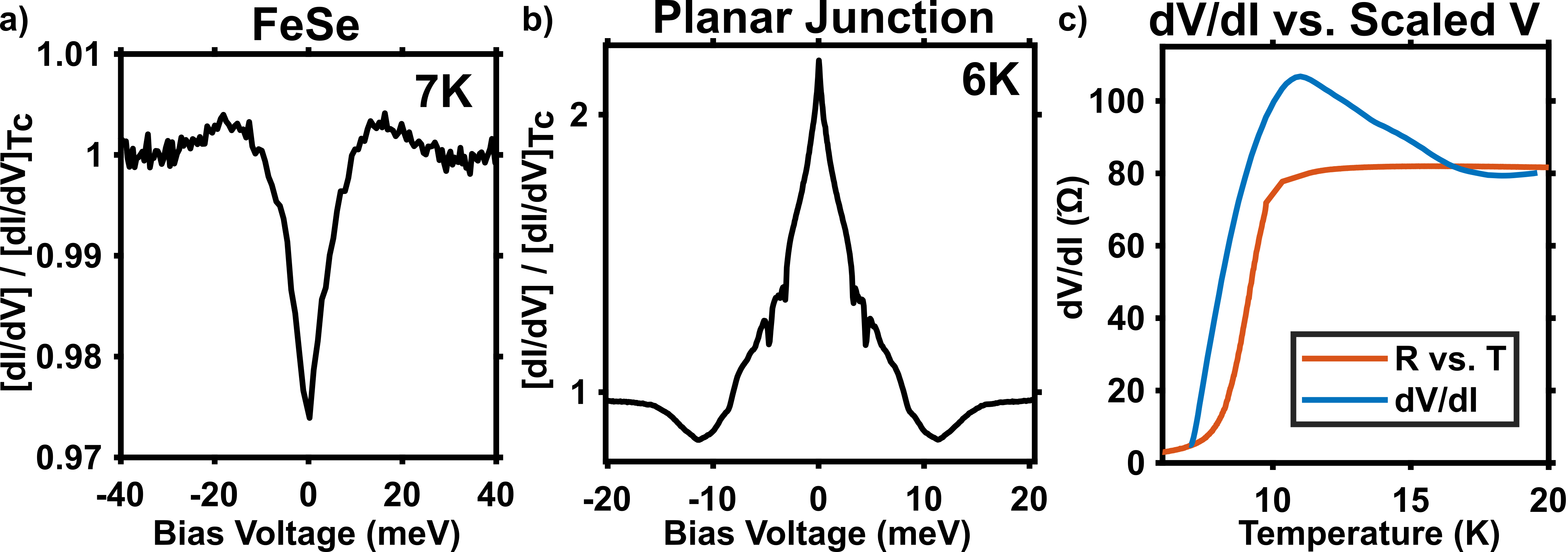}
    \caption{a) Soft-point contact on a bulk crystal of FeSe normalized to the critical temperature. b) Differential conductance using a planar junction, revealing a similar zero-bias peak. The smaller height results from the normal resistance of the Bi$_{2}$Te$_{2}$Se$_{1}$ that is in series with the tunnel contact. c) Differential resistance versus scaled voltage (blue) plotted along with the resistance versus temperature curve(orange). The strong overshoot of the voltage-dependent resistance and its return at high-bias to the normal state resistance confirms the spectra and zero bias conductance peak are not a result of heating.}
    \label{Controls}
\end{figure}

For additional confirmation that the ZBCP does not result from fabrication, exfoliation, impurities or the specific metal used in the contact, we performed a series of additional control experiments, summarized in Fig (\ref{Controls}). First, the topological gap in FTS closes with reduced tellurium levels, thus we expect the hinge mode is absent from FeSe. To confirm this as well as the irrelevance of contact type or normal metal used, we employed soft-point contact measurements. For FeSe we observe no evidence of an increase in conductance at zero bias below T$_{c}$ (see Fig (\ref{Controls}a). However, performing the same soft-point contact spectroscopy across multiple different FeTe$_{0.55}$Se$_{0.45}$ crystals always produces an increase in conductance at zero-bias when cooled below T$_{c}$ consistent with the data on contacts made via photolithography (see Supplemental Fig S3). The soft-point contacts revealed a smaller enhancement of the zero bias conductance in the superconducting state. However this is expected since the quasi-particle lifetime in the Ag paint contact is likely lower, which smears the spectra and reduces the height at zero bias. Similarly, we used planar junctions with Bi$_{2}$Te$_{2}$Se$_{1}$ via a method that has previously enabled spectroscopic studies with low barriers in van der Waals materials.\cite{Zareapour2012} As shown in Fig. \ref{Controls}b, these junctions also resulted in nearly identical spectra near zero bias. Here the lower zero bias conductance is expected as it contains contributions from the normal material being in series with the contact. Another extrinsic explanation for the peak is the interstitial Fe-atoms known to be present in these materials. However, we excluded this explanation by measurements on annealed samples where the Fe impurity content is dramatically reduced (see Supplemental Fig S3a), though the topology and Tc are only mildly affected.

An alternate mechanism for producing a ZBCP is Joule heating at the contact. We took a number of steps to rule this out. First, similar results were obtained regardless of the exact contact configuration (e.g. swapping contacts employed for current versus voltage in point contact or three-point measurements). In addition, we compared the voltage and temperature data by inverting the $\frac{dI}{dV}$ spectra and comparing it to the resistance versus temperature data taken on the same contact configuration (see Fig \ref{Controls}c). To align the two curves, we translate the $\frac{dV}{dI}$ curve such that zero voltage coincides with the temperature at which it was recorded (7 K). Next, we assume the voltage where the maximum resistance is measured is equivalent to heating to T$_{C}$, as this is the temperature where a peak in resistance is typically observed (see Fig \ref{HingeTheory}d). While the exact voltage dependence due to heating could be more complex, it is clear the $\frac{dV}{dI}$ versus voltage spectra are far in excess of the resistance measured at T$_{c}$, though at high bias they do return to the value measured at T$_{c}$. This further excludes voltage induced heating as the origin of the zero bias conductance peak. In addition, the background conductances in the c-axis, hinge, and point contacts are nearly identical. Therefore the heating across all of them should be approximately the same. However, they reveal quite distinct spectra (i.e. strong ZBCP in the hinge contact vs. nearly none in the c-axis) which, combined with the emergence of the zero-bias conductance peak (ZBCP) at T$_{c}$ in numerous contacts (see Figure \ref{MainResults} and Supplemental Figure S2), eliminates heating.
\par
In summary, via a variety of contact methods, we reveal helical hinge zero modes in the topological superconductor FeTe$_{0.55}$Se$_{0.45}$. Specifically,  contacts to the [001] surface made using hBN reveal standard Andreev reflection, while those draped over the hinge contain a cusp-like, zero-energy feature in the differential conductance. By combining with measurements using soft-point contacts on various crystals, we further confirm the intrinsic nature of this new mode. Furthermore, the appearance of an HHZM in FTS helps to establish both the topological and s$^{\pm}$ nature of the superconductivity. An important question raised by these results is the large size and the temperature dependence of the HHZM. It is possible that the large ratio of contact area to coherence length at the measured temperature ($\approx 1000x$), makes the measurement essentially many point-like contacts in parallel, leading to an apparently large conductance. The contact size may also play a role in the temperature dependence, as could the unknown size of the superconducting gap on the side surface. Thus future theoretical and experimental efforts must be made to better separate out the contact effects from the intrinsic response of the hinge mode we observe. 

\section{Acknowledgments}
K.S.B., R.O., and M.J.G. acknowledge support from the National Science Foundation, Award No. DMR-1709987.
Work at Brookhaven National Laboratory was supported by the Office of Science, U.S. Department of Energy under Contract No. DE-SC0012704.
Growth of hexagonal boron nitride crystals was supported by the Elemental Strategy Initiative conducted by the MEXT, Japan and the CREST (JPMJCR15F3), JST. The work at CCES is supported by  the Institute of Basic Science (IBS) in Korea (Grants No.IBS-R009-G2). We are grateful for numerous discussions with V. Yakovenko, V. Galitski, K.T. Law, R.-X. Zhang, W. Cole, K. Jiang and Z. Wang.

ASSOCIATED CONTENT: Supporting information available. Supporting information includes details regarding: exfoliation and fabrication of devices, experimental measurement setup, additional crystal measurements, and additional controls and checks performed on the devices.

%% Bibliography %%
%\bibliographystyle{achemso}
%\bibliography{Main/Round3/ZBCPinFTS}

\providecommand{\latin}[1]{#1}
\makeatletter
\providecommand{\doi}
  {\begingroup\let\do\@makeother\dospecials
  \catcode`\{=1 \catcode`\}=2 \doi@aux}
\providecommand{\doi@aux}[1]{\endgroup\texttt{#1}}
\makeatother
\providecommand*\mcitethebibliography{\thebibliography}
\csname @ifundefined\endcsname{endmcitethebibliography}
  {\let\endmcitethebibliography\endthebibliography}{}

\end{document}

\section{Experimental Details}
\subsection{Device Fabrication}
Single crystals of FeTe$_{0.55}$Se$_{0.45}$, provided by G. D. Gu, are freshly-cleaved or exfoliated in an inert argon glovebox. The newly exposed surfaces are then characterized via Raman Spectroscopy and Atomic Force Microscopy (AFM). Raman spectra show that the exfoliation process does not affect the surface quality of the crystal, while AFM data is used to confirm the thickness of each exfoliated flake. We made three different types of contact to the crystals: soft point-contacts, planar junctions with other single crystals, and mesoscale contacts with Cr/Au. Soft point-contacts were made by bending a copper wire to be in close proximity to the crystal, then touching a drop of heavily-diluted silver paint to the copper-crystal junction. The construction of the planar junctions is described by Zareapour P., \textit{et} \textit{al.} \cite{Zareapour2012}. To isolate the hinge of the crystal from electrical contact, we use an in-house built transfer stage and standard dry transfer techniques to cover the edge of the flake with hBN crystals provided by Takashi Tanaguchi and Kenji Watanabe. Patterns for mesoscale contacts were defined using standard photolithography techniques and our \textit{in situ} $\mu$PG101 direct-write lithography system. Contact areas are then cleaned with an argon plasma at high vacuum immediately before thermal deposition of 5nm Cr and 45nm of Au. Liftoff of metallic contacts is performed via light physical agitation in Microchem Remover PG at 50$^{o}$C.
\begin{figure}
    \centering
    \includegraphics[]{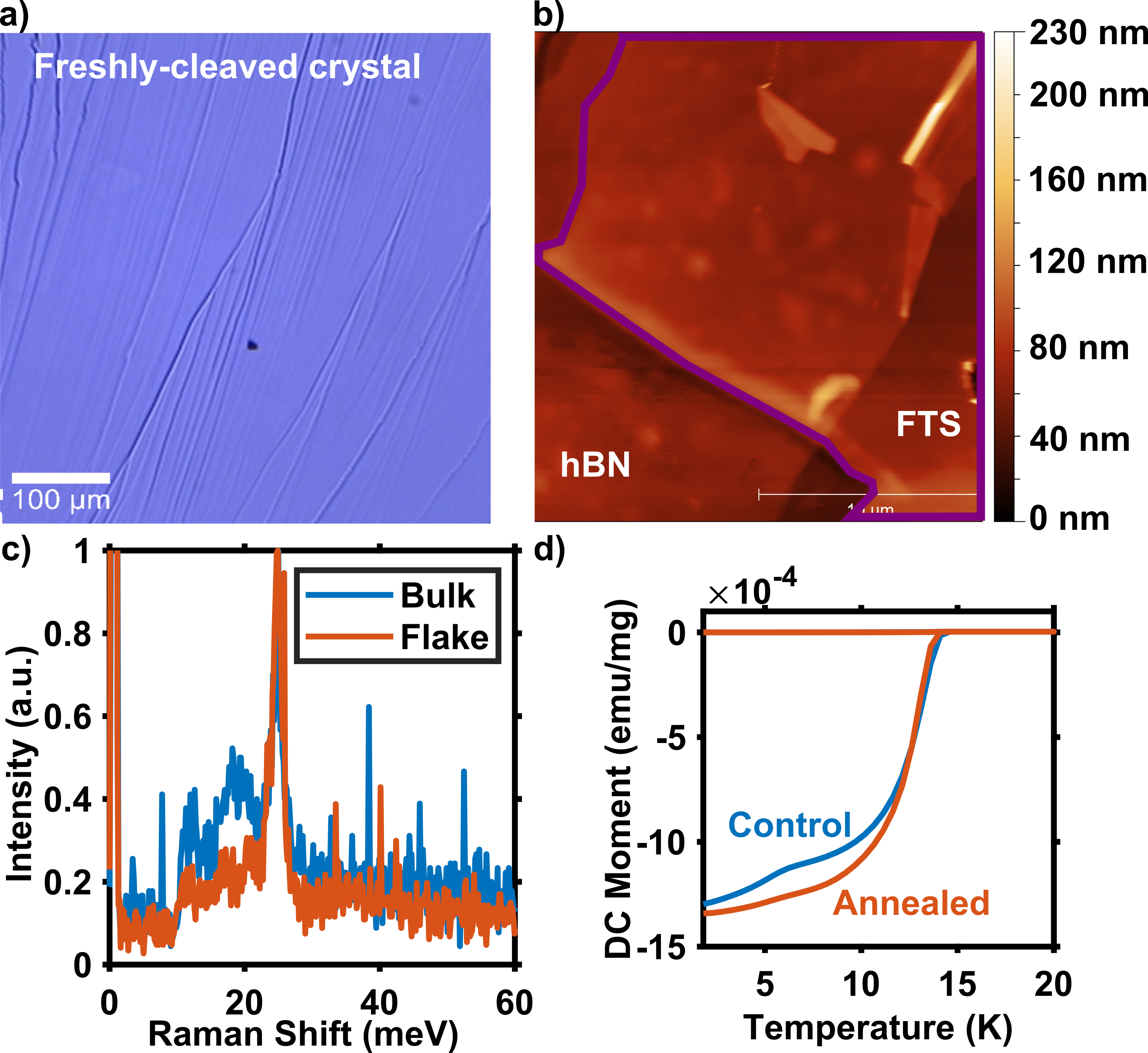}
    \caption{a) Optical Image of a freshly-cleaved crystal clearly showing many terraces. b) AFM image of the device shown in the main text. On the left is the hBN flake and the FTS flake is outlined in purple. c) Raman response of the FTS bulk crystal (blue) and exfoliated flake (red). d) DC moment of the bulk crystals before and after annealing. Annealing the crystal causes the superfluid density to increase.}
    \label{Device Characterization}
\end{figure}
\par
In order to test the effect of Fe-impurities on the result of the main text, we performed soft-point contact spectroscopy on both "as-received" crystals as well as crystals that were annealed in an Argon/Oxygen furnace at 400$^{o}$C. We expect that as the Fe-impurities are annealed out of the crystal, the superfluid density should increase. Indeed this is the case as can be seen in temperature dependent magnetic moment measurements (see Fig S1d). However, after annealing, we still observe a large ZBCP in the differential conductance eliminating the Fe-impurities as the origin of this zero-energy mode.

\subsection{Measurement Setup}
Differential conductance was measured directly with standard lock-in amplifier techniques. The SR810 Lock-In Amplifier is used to output an AC sine wave at 586.3Hz, while a BK Precision 1718b power supply is used to source the DC bias voltage. We use an in-house built adder box to first divide each voltage component down to appropriate levels, then add the two signals together which produces a DC voltage bias that has a 300$\mu$V AC voltage riding on top. The voltage across the device is measured directly with an Agilent multimeter for the DC component and an SR810 Lock-In Amplifier for the AC component. We place a known resistor in series with the device and measure both AC and DC voltages across the resistor to obtain the respective currents.

\begin{figure}
    \centering
    \includegraphics[width=\textwidth]{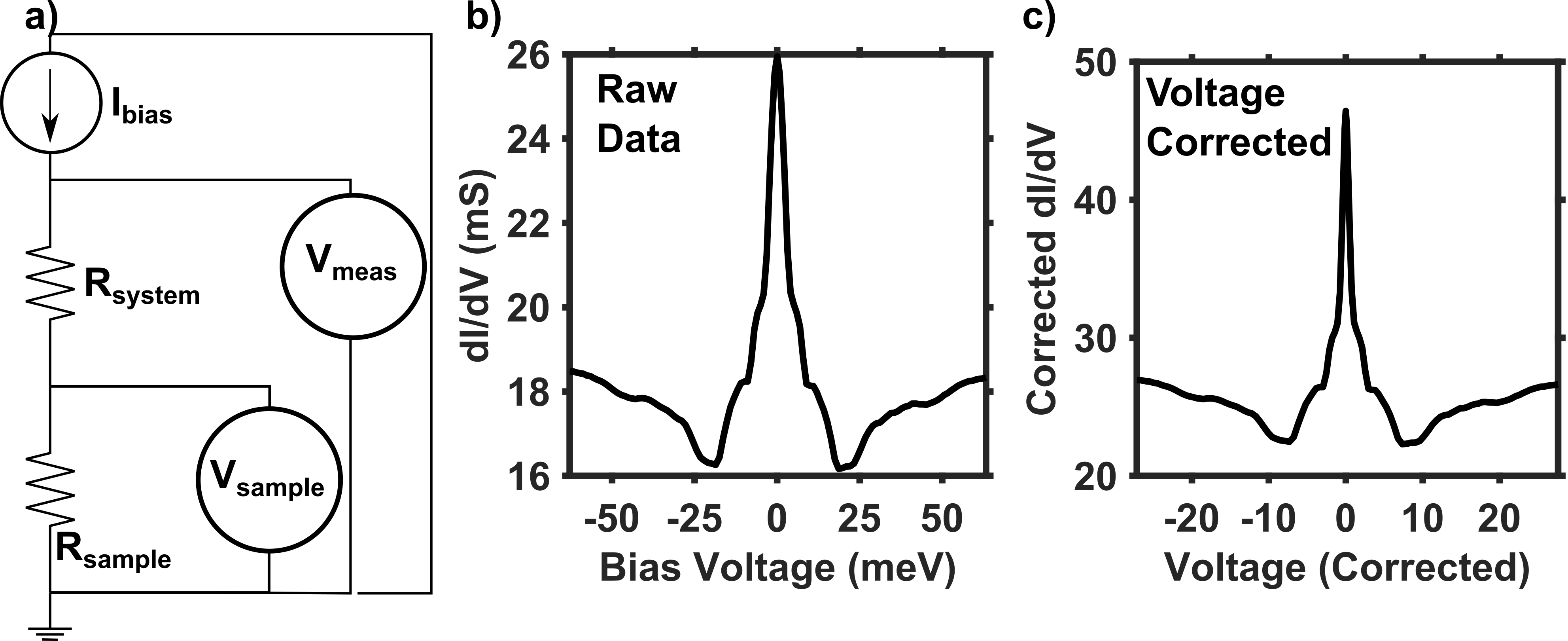}
    \caption{a) Circuit diagram showing the need for voltage correction. b) and c) Spectra before and after the voltage correction showing that while the relative magnitudes of the voltage and conductance change, the overall shape and features do not.}
    \label{Voltage Correction}
\end{figure}
\section{Voltage Correction}
It should be noted that the soft-point contacts require a correction to the measured voltage, as explained in Figure S\ref{Voltage Correction}. Since we do not break out the current and voltage contacts immediately after the junction, there is an additional voltage measured (See Figure S3a). To remove this extra voltage, we measure $R_{system}$ from the breakout to the junction and correct the voltages via:
\begin{align}
    V_{sample} = V_{measured}-I_{measured}*R_{system}
\end{align}
This voltage correction is not done to the mesoscale device presented in the main text, as we have separated the voltage and current contacts out close to the junction.

\begin{figure}[H]
    \centering
    \includegraphics[width=\textwidth]{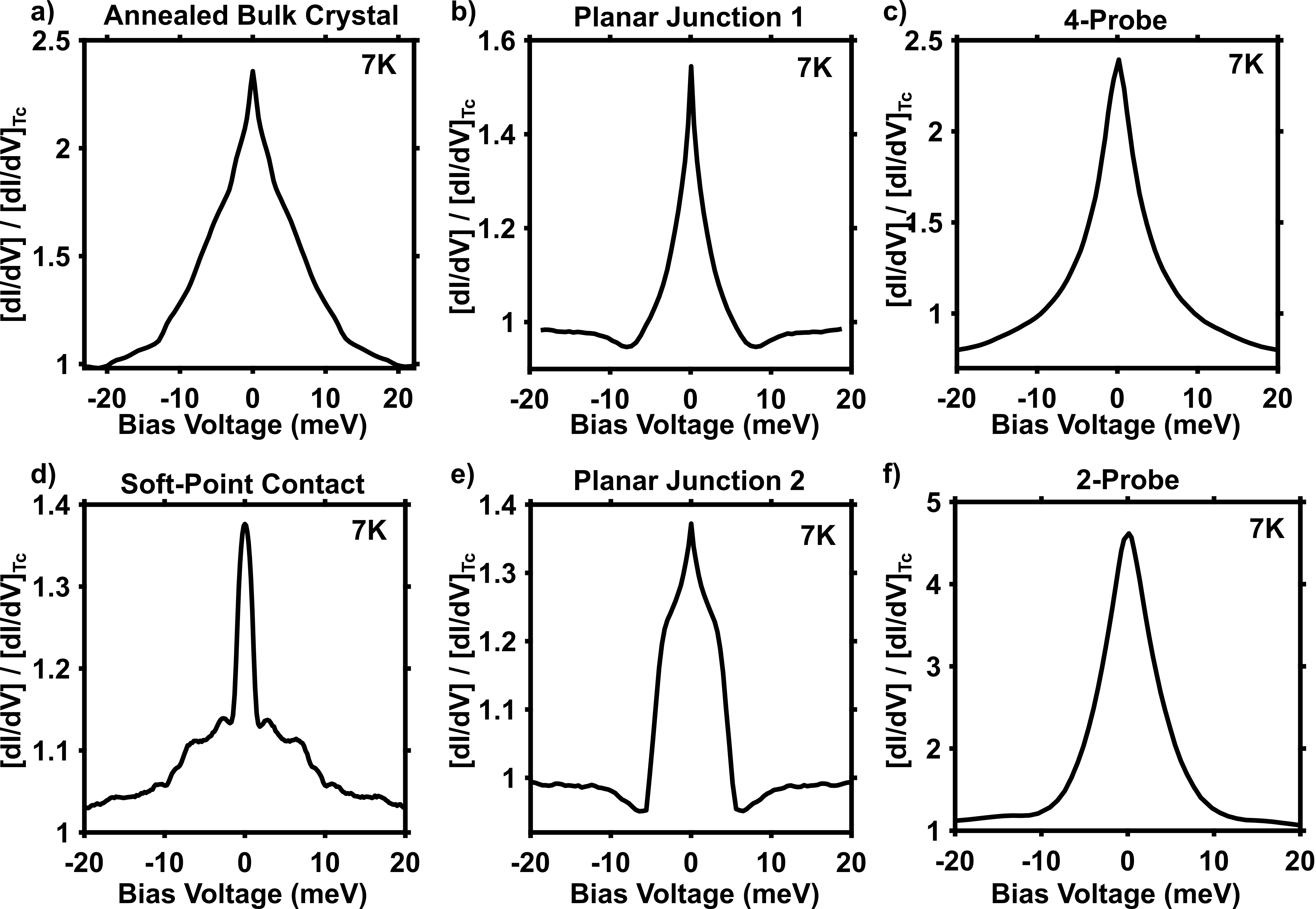}
    \caption{Measurements showing the robustness of the ZBCP across many different crystals and junctions types: a) Soft-point contact measurements of an Ar/H annealed bulk crystal. b) and e) Differential conductance across two planar junctions. c) Differential conductance using a 4-probe setup. d) Soft-point contact measurements on a freshly-cleaved bulk crystal. f) Differential conductance using a 2-probe setup with wide contacts.}
    \label{Robust ZBCP}
\end{figure}
\section{Reproducible ZBCP}
The NS junction data was reproduced across many separate bulk crystals and devices. Figure S2 shows six an additional six differential conductance spectra taken on various crystals that exhibit a ZBCP. As described in the fabrication details section, these spectra also represent three different types of junctions as well: lithographically-defined NS junctions (Fig S3c and f), soft-point contact junctions (Fig S3a and d), and planar junctions (Fig S3b and e). Although there are additional features on many of these spectra, the ZBCP and its response to temperature (i.e. quenching at critical temperature) are robust across all of our measurements.

\begin{figure}
    \centering
    \includegraphics[width=\textwidth]{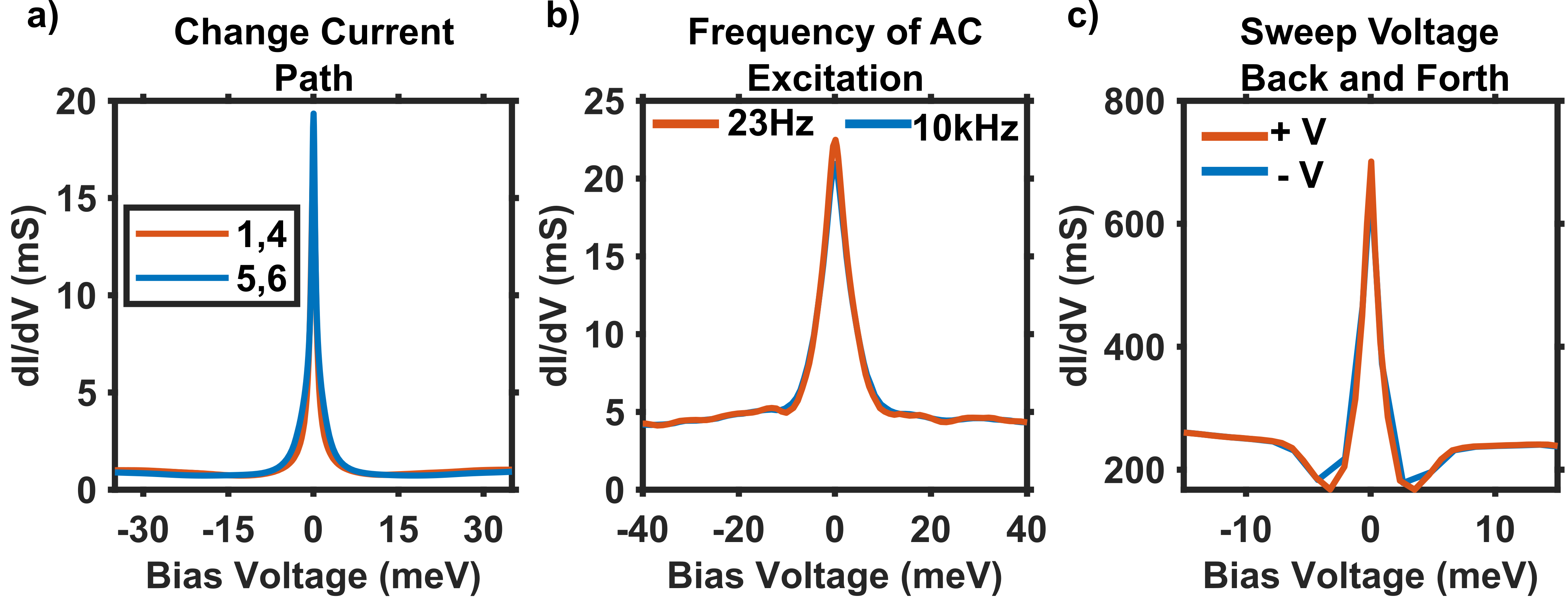}
    \caption{Additional systemic checks. a) Differential conductance performed on the same hinge contact but changing the current path. b) Differential conductance for an AC excitation of 23Hz and 10kHz, showing there's no difference for the two frequencies. c) Differential conductance spectra where the DC bias voltage is swept from positive to negative, then subsequently swept from negative to positive.}
    \label{Additional Checks}
\end{figure}
\section{Additional Measurement Checks}
As mentioned in the main text, we performed additional checks to be sure that our results presented in the main text are not due to systemic errors. One such error may be that we are simply putting too much current through the sample, causing the junction to heat up as we go to higher and higher voltages. If this were the case, then we would expect a transient resistance change to the sample, which should show up as anisotropy in the differential conductance when we sweep the voltage back and forth. However, when we check the voltage sweeps, we see no such transient response (See Figure S\ref{Additional Checks}c), which helps to eliminate Joule Heating as the main cause of the ZBCP. Next, we check that the current path through the device does not affect the ZBCP. As expected, switching the current and voltage leads for both source and drain leads does not affect the data (see Fig S4a).

%\bibliography{Main/Round2/ZBCPinFTS}

\providecommand{\latin}[1]{#1}
\makeatletter
\providecommand{\doi}
  {\begingroup\let\do\@makeother\dospecials
  \catcode`\{=1 \catcode`\}=2 \doi@aux}
\providecommand{\doi@aux}[1]{\endgroup\texttt{#1}}
\makeatother
\providecommand*\mcitethebibliography{\thebibliography}
\csname @ifundefined\endcsname{endmcitethebibliography}
  {\let\endmcitethebibliography\endthebibliography}{}